\documentclass[aps,prl,twocolumn,superscriptaddress,amsmath]{revtex4}
\usepackage{multirow} 
\usepackage{graphicx}% Include figure files
\usepackage{psfrag}

\newcounter{saveeqn}
\newcommand{\alpheqn}{\setcounter{saveeqn}{\value{equation}}
\stepcounter{saveeqn}
\setcounter{equation}{0}\renewcommand{\theequation}{\mbox{\arabic{saveeqn}\alph{equation}}}}
\newcommand{\reseteqn}{\setcounter{equation}{\value{saveeqn}}
\renewcommand{\theequation}{\arabic{equation}}}

\begin{document}

\title{Sampling rare switching events in biochemical networks}

\author{Rosalind J. Allen}

\affiliation{FOM Institute for Atomic and Molecular Physics, Kruislaan 407, 1098 SJ Amsterdam, The Netherlands}

\author{Patrick B. Warren}

\affiliation{Unilever R \& D Port Sunlight, Bebington, Wirral CH63 3JW, United Kingdom}

\author{Pieter Rein ten Wolde}
 
\email{tenwolde@amolf.nl}
 
\affiliation{FOM Institute for Atomic and Molecular Physics, Kruislaan 407, 1098 SJ Amsterdam, The Netherlands}

\date{\today}

\begin{abstract}
Bistable biochemical switches are ubiquitous in gene regulatory
networks and signal transduction pathways. Their switching dynamics,
however, are difficult to study directly in experiments or
conventional computer simulations, because switching events are rapid,
yet infrequent. We present a simulation technique that makes it
possible to predict the rate and mechanism of flipping of biochemical
switches. The method uses a series of interfaces in phase space
between the two stable steady states of the switch to generate
transition trajectories in a ratchet-like manner. We demonstrate its
use by calculating the spontaneous flipping rate of a symmetric model
of a genetic switch consisting of two mutually repressing genes. The
rate constant can be obtained orders of magnitude more efficiently
than using brute-force simulations. For this model switch, we show
that the switching mechanism, and consequently the switching rate,
depends crucially on whether the binding of one regulatory protein to
the DNA excludes the binding of the other one.  Our technique could
also be used to study rare events and non-equilibrium processes in
soft condensed matter systems.
\end{abstract}

\maketitle

Biochemical switches are essential for the functioning of living cells. These switches are
networks of chemical reactions that exhibit more than one stable
steady state; in the presence of noise, flipping can occur between these states.  Well-characterized examples include the lysis-lysogeny
switch in bacteriophage $\lambda$ \cite{ptashne} and the lac repressor
in {\em{E. Coli}} \cite{muller,oudenaarden,vilar}. Experimental and
theoretical studies have established the presence of bistability in
other biochemical networks, including those regulating the cell cycle
and developmental fate
\cite{pomerening,sha,ferrell1,xiong,angeli,ferrell2}. In addition,
synthetic switches have been constructed {\em{in vivo}}
\cite{gardner,atkinson,becskei1}.

 Computational modeling has an important role to play in explaining
the properties of biochemical switches. A  stochastic approach is required to obtain  the mechanism and rate of switching,  since these
switches are flipped by noise. Examples of such approaches are the
chemical Langevin technique~\cite{vankampen}, analysis of the chemical master equation \cite{gardiner} or stochastic simulation techniques.   Simulation algorithms that
generate trajectories consistent with the chemical master equation
include the Gillespie Algorithm \cite{gillespie1,gillespie} and
StochSim \cite{stochsim}. Where spatial resolution is required,
methods such as Green's Function Reaction Dynamics can be used
\cite{gfrd}.

Biochemical switches are often very difficult or impossible to simulate
using the above techniques in a brute-force manner. This is because
they can be extremely stable, showing few or no flips during the
accessible simulation time. The average number of spontaneous
transitions from the lysogenic to the lytic state for bacteriophage
$\lambda$, for example, is about one in $10^7$ bacterial generations
\cite{aurell,aurell2}.  New methods are therefore required to model such important but rare
events in biochemical networks.

Techniques for the simulation of rare events have been developed in
the field of soft condensed matter physics \cite{daan}. Recent
developments focus on the transition path ensemble (TPE). For a rare
transition between stable states $A$ and $B$, this is the set of all
`reactive' trajectories leading from $A$ to $B$ (transition paths).  Analysis of
the TPE gives detailed information on the transition mechanism and
leads to a prediction of the rate constant. Transition Path Sampling
(TPS) methods have been developed to generate members of this ensemble
in a computationally efficient way \cite{dellago,tps}. TPS has been applied to a wide variety of
problems, including chemical reactions in solution, conformational
transitions in biopolymers and protein folding \cite{bolhuis}.

Biochemical switches, however, differ fundamentally from these
problems. As we shall discuss, in simulations of reaction networks the
stationary distribution of states is generally not known {\em a priori}. As a
result, TPS methods cannot straightforwardly be applied.

In this article, we present a new scheme 
for sampling the TPE and computing the rate constant. This ``Forward
Flux Sampling'' (FFS) method is efficient and straightforward. It does not require prior knowledge of the
phase-space density and can be applied to simulations of
biochemical networks. The method could also be implemented in any
other stochastic dynamics scheme. To our knowledge, FFS constitutes a novel approach to
sampling the TPE. Rather than generating transition paths one at a
time (as in TPS), a large number of paths are grown simultaneously from state $A$
to state $B$ in a series of connected layers.

As an application of the FFS method, we have calculated the
spontaneous flipping rate of a simple genetic switch, consisting of
two mutually repressing genes. We show, in agreement with previous
work \cite{warren1}, that the stability of this switch is greatly
enhanced when the operator regions for the two genes are
 mutually exclusive, and that this is due to an important change in the
flipping mechanism.

\section*{Background}

In this article, the FFS method is used to calculate switching rates
for biochemical networks simulated with the Gillespie Algorithm
\cite{gillespie1,gillespie}.  This algorithm is an application to
chemical reactions of the kinetic Monte Carlo technique
\cite{barkema}, first introduced by Bortz {\em{et al}}
\cite{bortz}. The system is specified by a set of chemical components
$\{X\}$ and a list of allowed reactions, together with their rate
constants. The concentrations $\{n_X\}$ of all the components are
assumed to be homogeneous in space; the state of the system at any
instant in time is defined by $\{n_X(t)\}$. The concentrations
$\{n_X(t)\}$ are propagated stochastically in time, assuming each
reaction to be a Poisson process. This time propagation is consistent
with the chemical master equation, so that a Gillespie simulation is
in fact a numerical solution of the master equation.

  An important feature of the Gillespie Algorithm, and of other
  methods for simulating reaction networks, is that the distribution
  of states, {\em i.e.} the phase space density, is not known {\em{a
  priori}}, but is an output of the simulation. The phase space
  density can be obtained by solving the chemical master equation
  \cite{reichl}, but this is generally a demanding task, which is
  indeed often the motivation for carrying out a Gillespie
  simulation.

Transition Path Sampling (TPS) has been developed to study rare events
in condensed-matter systems. In TPS, paths belonging to the TPE are
obtained by importance sampling in trajectory space. New paths
connecting stable states $A$ and $B$ are generated by making changes
to existing paths. A new path is accepted or rejected according to its
weight in the TPE, which depends on the phase space density of its
initial point, as well as the transition probability for each
subsequent step. Without prior knowledge of the stationary
distribution of states, however, this approach cannot conveniently be applied.

The FFS algorithm which we present in this article differs
fundamentally from these methods. Rather than generating transition
paths one at a time, many paths are grown simultaneously, in a series
of layers, each of which forms the basis for the next one. Prior
knowledge of the stationary distribution of states is not
required. FFS is well suited for convenient and efficient calculation
of switching rates in biochemical reaction networks. The FFS method is not limited to reaction networks: although it cannot be used for systems with deterministic dynamics, it is applicable to any stochastic dynamical scheme, such as Langevin or Brownian Dynamics. In this context, it could be used to study rare events in soft condensed matter systems such as protein folding and crystal nucleation, or non-equilibrium processes such as DNA or RNA stretching.

\section*{Rate Expression}

The expression for the rate constant that is used in the FFS algorithm is the same as that described by van Erp {\em{et al}} \cite{vanerp}. The transition occurs between two phase space regions  $A$ and $B$, which must both be ``stable'' in the sense that if the the system is placed outside these regions, it will rapidly evolve in time towards one of them. $A$ and $B$  are characterized by the functions  $h_A(x)$ and $h_B(x)$ (where $x$ denotes all co-ordinates of the phase space: in the case of the Gillespie algorithm the concentrations of all the system components), such that:
\begin{eqnarray}
h_A(x)=1\,\, {\rm{if}}\,\, x \in A,\,\, {\rm{else}}\,\, h_A(x)=0\\\nonumber
h_B(x)=1\,\, {\rm{if}}\,\, x \in B,\,\, {\rm{else}}\,\, h_B(x)=0
\end{eqnarray}
We also define the functions  $h_{\mathcal{A}}$ and  $h_{\mathcal{B}}$, which depend not only on $x(t)$ but also on the history of the system: $h_{\mathcal{A}}=1$ if the system was more recently in $A$ than in $B$, and is zero otherwise, while $h_{\mathcal{B}}=1$ if the system was more recently in $B$ than in $A$, and is zero otherwise. Thus $h_{\mathcal{A}} + h_{\mathcal{B}}=1$ for any point on any path in phase space.

The rate constant $k_{AB}$ for transitions from region $A$ to region $B$ is given by:
\begin{equation}\label{eq3}
k_{AB} = \frac{{\overline{\Phi}}_{A,B}}{{\overline{h}}_{\mathcal{A}}}
\end{equation}
Here, ${\overline{\Phi}}_{A,B}$ is the average number of trajectories per unit time entering region $B$, coming directly from $A$ ({\em{i.e.}} which were in $A$ more recently than they were in $B$).  Here, the overbar denotes an average over all phase space points, with their associated histories.

The flux ${\overline{\Phi}}_{A,B}$ in Equation (\ref{eq3}) is difficult to obtain accurately from a simulation because the system makes few, if any, spontaneous crossings from $A$ to $B$ in a typical run. To alleviate this problem, a parameter $\lambda$ is chosen, such that the functions $h_A$ and $h_B$ can be written as: 
\begin{eqnarray}
h_A(x)=1\,\, {\rm{if}}\,\, \lambda (x) \le \lambda_A,\,\, {\rm{else}}\,\, h_A(x)=0\\\nonumber
h_B(x)=1\,\, {\rm{if}}\,\, \lambda (x) \ge \lambda_B,\,\, {\rm{else}}\,\, h_B(x)=0
\end{eqnarray}
An increasing  series of values of $\lambda$, $\{\lambda_1 \dots \lambda_n\}$, is then chosen, such that $\lambda_1 \ge \lambda_A$ and $\lambda_n < \lambda_B$. These must constitute  non-intersecting surfaces in phase space. It is not necessary for $\lambda$ to be the reaction coordinate, merely that $\lambda_A$ and $\lambda_B$ describe the two stable states (the exact positioning of these surfaces is not critical). Moreover, the system should not reach any $\lambda_{i+1}$ before it has crossed the preceding surface $\lambda_i$. Defining $P(\lambda_{i+1}|\lambda_{i})$ as the probability that a trajectory which passes through $\lambda_{i}$ coming directly from $A$ ({\em{i.e.}} having been in $A$ more recently than it last crossed $\lambda_i$), will subsequently reach the surface $\lambda_{i+1}$ before returning to $A$, equation (\ref{eq3}) can be written as:
\begin{equation}\label{eq6}
k_{AB} = \frac{{\overline{\Phi}}_{A,B}}{{\overline{h}}_{\mathcal{A}}} = \frac{{\overline{\Phi}}_{A,1}}{{\overline{h}}_{\mathcal{A}}} P(\lambda_B|\lambda_1)
\end{equation}
Expression (\ref{eq6}) indicates that the total flux from $A$ to $B$ is simply the total flux from $A$ to $\lambda_1$, multiplied by the probability that a trajectory reaching $\lambda_1$ from $A$ will eventually arrive in $B$, before returning to $A$. A key point is that $P(\lambda_B|\lambda_1)$ can be expressed as the product of the probabilities of reaching each successive interface from the previous one, without returning to $A$:
\begin{equation}\label{eq7}
P(\lambda_B|\lambda_1) = \prod_{i=1}^{n-1}P(\lambda_{i+1}|\lambda_{i}) \times P(\lambda_B|\lambda_n)
\end{equation}
so that 
\begin{equation}\label{eq8}
-\log{P(\lambda_B|\lambda_1)} = -\sum_{i=1}^{n-1}\log{P(\lambda_{i+1}|\lambda_{i})} -\log{P(\lambda_B|\lambda_n)}
\end{equation}
Expressions (\ref{eq3}) and (\ref{eq6})-(\ref{eq8}) are used in the FFS method to calculate $k_{AB}$.

\section*{Forward Flux Sampling}
The first stage of the FFS algorithm involves the choice of the parameter $\lambda$ and values for $\lambda_A$, $\lambda_B$ and $\{\lambda_1 \dots \lambda_n\}$.  In any one chemical reaction step, the system must be able to cross at most one surface $\lambda_i$. Of course, some choices of $\lambda$ will lead to more efficient path sampling than others, but we shall demonstrate in the next sections that for a typical genetic switch, a rather simple definition of $\lambda$ gives very satisfactory results. It is also convenient to define a series of ``sub-surfaces'' $\{\lambda_i^{(1)} \dots \lambda_i^{(m_i)}\}$, in between each pair of surfaces $\lambda_i$ and $\lambda_{i+1}$, such that $\lambda_i^{(1)}=\lambda_i$ and $\lambda_i^{(m_i)}=\lambda_{i+1}$.

\begin{figure}
\begin{center}
{\rotatebox{0}{{\includegraphics[scale=0.7,clip=true]{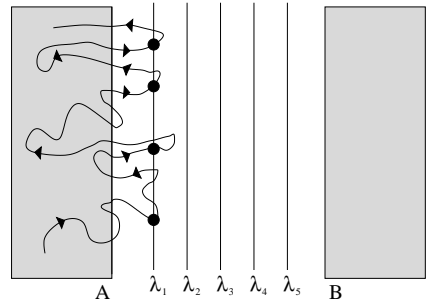}}}}
\caption{The first stage of the FFS method.\label{fig1}}
\end{center}
\end{figure}

A simulation is then carried out starting from a point in region $A$. After an equilibration period, the value of $\lambda$ is monitored during a run of length $T$. Whenever the trajectory crosses the surface $\lambda_1$, coming directly from $A$, a counter $N_f$ is incremented. If $N_f$ is less than a user-defined number $C_1$, the phase space co-ordinates of the system are also stored. The run is then continued. After simulation time $T$,  one is left with a collection of  $C_1$ points at or just beyond $\lambda_1$, as well as a measurement of the flux ${\overline{\Phi}}_{A,1}/{\overline{h}}_{\mathcal{A}}= N_f/T$.  This procedure is illustrated schematically in Figure \ref{fig1}: crossings of surface $\lambda_1$ that are labeled with a black circle contribute to $N_f$ and to the collection of points at $\lambda_1$. \footnote{If the system happens to enter region $B$ during this run, it is replaced in $A$, re-equilibrated, and the run continued.}

\begin{figure}
\begin{center}
{\rotatebox{0}{{\includegraphics[scale=0.7,clip=true]{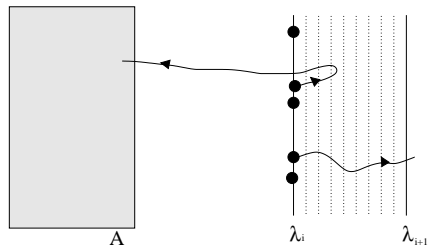}}}}
\caption{The second stage of the FFS method.\label{fig2}}
\end{center}
\end{figure}

Figure \ref{fig2} illustrates the next stage of the algorithm. The collection of points at $\lambda_1$ is used to initiate a large number $M_1$ of short simulation trial runs. In each of these trials, a phase space point from the collection at $\lambda_1$ is chosen at random. This is then used as the starting point for a simulation run, which is continued until the system crosses either $\lambda_2$ or $\lambda_A$. During this run, the maximum value of $\lambda$, $\lambda_{max}$,  achieved by the system is recorded. Counters $N_{1}^{j}$ for all the sub-surfaces $\lambda_1^{(j)} \le \lambda_{max}$ are then incremented by one.   After $M_1$ trials, a good estimate is obtained for $P(\lambda_{1}^{(j)}|\lambda_1) = N_{1}^{j}/M_1$, for $1 \le j \le m_1$. We note that $P(\lambda_{1}^{(1)}|\lambda_1)=1$ and  $P(\lambda_{1}^{(m_1)}|\lambda_1)=P(\lambda_{2}|\lambda_1)$.
 
During the trial procedure outlined above, one also makes a new collection of $C_2$ points at or just beyond the surface $\lambda_2$: these are the final phase space points of those trial runs starting from $\lambda_1$ which make it to $\lambda_2$. The number $M_2$ of trials must be large enough to generate $C_2$ points at $\lambda_2$. The values of $M_2$, $C_2$ and $M_i$ and $C_i$ for all the subsequent surfaces $2 \le i \le n$ are chosen by the user: the $C_i$ should be large enough to allow good sampling of the phase space.

The trial run procedure is repeated for each subsequent surface $\lambda_i$, starting from the collection of $C_i$ phase space points generated by the successful runs from $\lambda_{i-1}$. Eventually $\lambda_B$ is reached, and one is left with a series of histograms $P(\lambda_{i+1}^{(j)}|\lambda_i)$, for $1 \le i \le n$ and $1 < j \le m_i$. Using equation  (\ref{eq8}), these histograms can be fitted together to obtain a smooth curve $P(\lambda|\lambda_1)$ \cite{ferrenberg,duijneveld}, the value of which at $\lambda=\lambda_B$ is $P(\lambda_B|\lambda_1)$. The rate constant $k_{AB}$ is obtained on multiplying $P(\lambda_B|\lambda_1)$ by the flux ${\overline{\Phi}}_{A,1}/{\overline{h}}_{\mathcal{A}}$ calculated in the first stage of the algorithm.

It is important to remark that the FFS algorithm does not assume that
 the distribution of phase space points at the interfaces $\{\lambda_1
 \dots \lambda_n\}$ is equal to the stationary
 distribution of states.  For the example which we present in this
 paper, this turns out to have significant consequences for the
 transition mechanism.

\section*{Application: A Genetic Switch}
We have applied the FFS method to a simplified model of a genetic
toggle switch \cite{warren1,kepler,cherry}. This model could be regarded as a
minimal representation of the lysis-lysogeny switch in bacteriophage
$\lambda$ \cite{ptashne}; a synthetic switch of this type has also
been constructed {\em{in vivo}} \cite{gardner}.

The model switch consists of two proteins $\mathrm{A}$ and
$\mathrm{B}$ and their corresponding genes $A$ and $B$.  $\mathrm{A}$
and $\mathrm{B}$ form homodimers $\mathrm{A}_2$ and $\mathrm{B}_2$
which can bind to the DNA strand (here labeled $\mathrm{O}$) and
influence transcription. When the dimer $\mathrm{A}_2$ is bound to the
DNA, gene $B$ is not transcribed, while $\mathrm{B}_2$, when bound,
correspondingly blocks transcription of gene $A$: thus $\mathrm{A}$
and $\mathrm{B}$ mutually repress one another's production. Both
proteins are also degraded in the monomer form. We consider two
versions of this switch: the ``general'' switch, in which both dimers
can bind simultaneously to the DNA, forming the species
$\mathrm{O}\mathrm{A}_2\mathrm{B}_2$, and the ``exclusive'' switch, in
which only one dimer can be bound at any time. The exclusive switch
models the case where the operator regions of genes $A$ and $B$ are
overlapping.

The switch is represented by the set of reactions  (\ref{eq9}).

\alpheqn

\begin{eqnarray}\label{eq9}
2\mathrm{A} \rightleftharpoons \mathrm{A}_2 & \qquad  & 2\mathrm{B} \rightleftharpoons \mathrm{B}_2\label{equationa}\\
\mathrm{O} + \mathrm{A}_2 \rightleftharpoons \mathrm{O}\mathrm{A}_2  & \qquad & \mathrm{O} + \mathrm{B}_2 \rightleftharpoons \mathrm{O}\mathrm{B}_2\label{equationb}\\
 \mathrm{O}\mathrm{A}_2 + \mathrm{B}_2 \rightleftharpoons \mathrm{O}\mathrm{A}_2\mathrm{B}_2  & \,\, * \,\, &\mathrm{O}\mathrm{B}_2 + \mathrm{A}_2 \rightleftharpoons \mathrm{O}\mathrm{A}_2\mathrm{B}_2 \qquad \label{equationc}\\
 \mathrm{O} \to \mathrm{O} + \mathrm{A}  &\qquad & \mathrm{O} \to \mathrm{O} + \mathrm{B}\label{equationd}\\
 \mathrm{O}\mathrm{A}_2 \to \mathrm{O}\mathrm{A}_2 + \mathrm{A}  &\qquad & \mathrm{O}\mathrm{B}_2 \to \mathrm{O}\mathrm{B}_2 + \mathrm{B}\label{equatione}\\
 \mathrm{A} \to \emptyset  &\qquad & \mathrm{B} \to \emptyset \label{equationf}
\end{eqnarray}

\reseteqn

The asterisk indicates that reaction (\ref{equationc}) happens only for the general switch. Here, we study a symmetrical version of the switch: the rate constants for the reactions on the left and right-hand sides of scheme (\ref{eq9}) are identical. These are all expressed in terms of the protein production rate constant $k$ (for reactions (\ref{equationd}) and (\ref{equatione})), so that the unit of time in our calculations is $k^{-1}$. The rate constants for both the forward and backward dimerization reactions (\ref{equationa}) are $5k$. Binding to the DNA occurs with rate constant $5k$ and dissociation of the complex with rate constant $k$ (reactions (\ref{equationb}) and (\ref{equationc})). Finally, the rate constant for protein degradation, reactions (\ref{equationf}), is $0.25k$. These parameters are chosen  such that in a simulation using the Gillespie algorithm, the switch flips between the $\mathrm{A}$- and $\mathrm{B}$-rich states  at a rate that can be measured by brute-force simulation. This allows us to test the FFS method.  The model is, of course, highly simplistic: our aim is here to  demonstrate  the FFS scheme using a simple example.

\section*{Results}
\begin{figure}
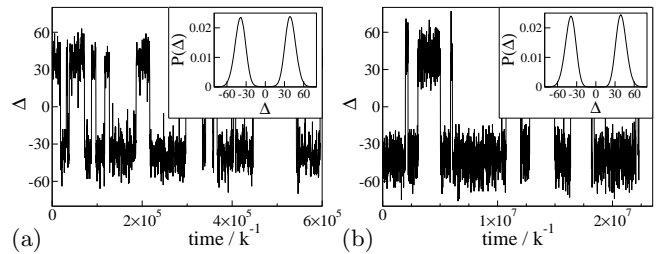

\begin{center}
\makebox[0pt][l]{(a)}{\rotatebox{0}{{\includegraphics[scale=0.175,clip=true]{fig3a}}}}\makebox[0pt][l]{(b)}{\rotatebox{0}{{\includegraphics[scale=0.175,clip=true]{fig3b}}}}
\caption{$\Delta$ as a function of time (in units of $k^{-1}$) for a
typical simulation run, for the general (a) and exclusive (b)
switches. Insets show probability $P(\Delta)$ of observing a
particular value of $\Delta$, calculated over a total simulation time
of $1 \times 10^7 k^{-1}$ (general switch, (a)) and $5 \times 10^9
k^{-1}$ (exclusive switch, (b)).\label{fig3} }
\end{center}
\end{figure}

Figure \ref{fig3} shows the results of Gillespie simulations of the model genetic switch, in the general (a) and exclusive (b) cases. The difference $\Delta$ in the total copy numbers of the two proteins, $\Delta = \mathrm{N_B}-\mathrm{N_A}$, is plotted as a function of time, where  $\mathrm{N_A}$ and $\mathrm{N_B}$ are defined by:
\begin{eqnarray}
\mathrm{N_A} = n_\mathrm{A} + 2n_{\mathrm{A}_2} + 2n_{\mathrm{O}\mathrm{A}_2} + 2n_{\mathrm{O}\mathrm{A}_2\mathrm{B}_2}\\\nonumber
\mathrm{N_B} = n_\mathrm{B} + 2n_{\mathrm{B}_2} + 2n_{\mathrm{O}\mathrm{B}_2} + 2n_{\mathrm{O}\mathrm{A}_2\mathrm{B}_2}
\end{eqnarray}
Of course, $n_{\mathrm{O}\mathrm{A}_2\mathrm{B}_2}=0$ for the exclusive switch. Noting the different scales on the time axis, the exclusive switch (b) has a much lower flipping rate than the general switch (a), in agreement with previous work \cite{warren1}. The probability $P(\Delta)$ of obtaining a particular value of $\Delta$ is shown in  the insets, demonstrating clearly that both the general and exclusive switches are bistable.

Using long brute-force Gillespie simulations, the rate constant for
the transition from the $\mathrm{A}$-rich to the $\mathrm{B}$-rich
state was obtained for each switch. We define phase space region $A$
to be where $\Delta \le -25$, and region $B$ to be where $\Delta \ge
25$. The system flips stochastically between the state where
$h_{\mathcal{A}}=1$ and $h_{\mathcal{B}}=0$ ({\em{i.e.}} it was most
recently in $A$) and that where $h_{\mathcal{A}}=0$ and
$h_{\mathcal{B}}=1$ (it was most recently in $B$). The times $t$
between flipping events are distributed according to a Poisson
distribution $p(t)=k_{AB}\exp{[-k_{AB}t]}$, where $k_{AB}=k_{BA}$
since the switch is symmetrical in $A$ and $B$. The rate constant
$k_{AB}$ can conveniently be measured by fitting the cumulative
distribution $F(t) = \int_0^{t} dt'\, p(t') $ to the function
$1-\exp{[-k_{AB}t]}$. This procedure resulted in values of
$k_{AB}=(4.21 \pm 0.05) \times 10^{-5} k$ for the general switch and
$k_{AB}=(9.4 \pm 0.2) \times 10^{-7} k$ for the exclusive switch
(using simulation runs of total length $3 \times 10^8 k^{-1}$ [12546
flips observed] and $9 \times 10^9 k^{-1}$ [8808 flips observed]
respectively).

We next re-calculated $k_{AB}$ using FFS. The surfaces $\{\lambda_1 \dots \lambda_n\}$, were defined in terms of $\Delta$: {\em{i.e.}} $\lambda=\Delta$. Regions $A$ and $B$ are given by $\lambda=\Delta \le \lambda_A$ and $\lambda=\Delta \ge \lambda_B$, respectively. To be sure that the exact values of $\lambda_A$, $\lambda_B$ and $\{\lambda_1 \dots \lambda_n\}$ did not affect the result, these parameters were varied in a series of separate calculations.  In all cases, we set $\lambda_1=\lambda_A$, and at each surface  $\lambda_i$, $C_i=10000$ points were stored and  $M_i=100000$ shooting trials were made. Results were averaged over $10$ independent calculations, to obtain error bars similar to those of the brute-force results.

\begin{figure}
\begin{center}
{\rotatebox{0}{{\includegraphics[scale=1,clip=true]{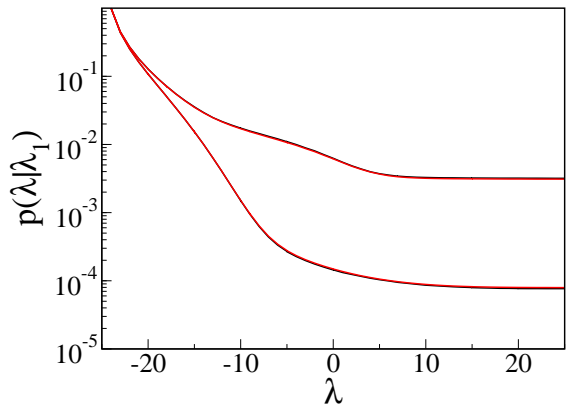}}}}
\caption{$P(\lambda|\lambda_1)$ for the general (top) and exclusive (bottom) switches, where $\lambda_A=-24$ and $\lambda_B=25$. Black lines: brute-force simulation results (averaged over a total time $1 \times 10^{8}k^{-1}$ for the general switch and $9 \times 10^{9}k^{-1}$ for the exclusive switch); Red lines: FFS results, averaged over 10 independent calculations. \label{fig4}}
\end{center}
\end{figure}

Figure \ref{fig4} shows $P(\lambda|\lambda_1)$ as a function of $\lambda$, for the general and exclusive switches. This function can also be obtained from an analysis of the brute-force simulation trajectories: the figure shows excellent agreement between the brute-force results  (shown in black), and  those of FFS (shown in red).

\begin{table}[t]
\begin{center}
\begin{tabular}{ccccc}
\hline
\multicolumn{5}{l}{General switch}\\
%\hline 
\,\,$\lambda_B$\,\, & \,\,$n$ \,\,&\,\, $f / k \times 10^{-2}$\,\, &\,\, $P(\lambda_B|\lambda_1) \times 10^{-3}$\,\, &\,\, $k_{AB} / k \times 10^{-5}$\,\,\\
$30$  & $14$ & $2.97 \pm 0.01$ & $1.41 \pm 0.03$ & $4.19 \pm 0.07$\\
$25$  & $11$ & $1.33 \pm 0.01$ & $3.10 \pm 0.06$ & $4.11 \pm 0.07$\\
$20$ & $9$ & $0.392\pm 0.003$ & $10.5 \pm 0.1$ & $4.13 \pm 0.04$ \\
$\bf 25$  & - & $\bf 1.339 \pm 0.004$ & $\bf 3.15 \pm 0.05$ & $\bf 4.22 \pm 0.06$\\
  &  & & & $\bf 4.21 \pm 0.05$\\
%\hline
\multicolumn{5}{l}{Exclusive switch}\\
%\hline 
\,\,$\lambda_B$\,\, &\,\, $n$\,\, &\,\, $f / k \times 10^{-2}$\,\, &\,\, $P(\lambda_B|\lambda_1) \times 10^{-5}$\,\, &\,\, $k_{AB} / k \times 10^{-7}$\,\,\\
$30$ & $16$ & $2.98 \pm 0.01$ & $3.2 \pm 0.1$ & $9.5 \pm 0.3$\\
$25$ & $11$ & $1.211 \pm 0.007$ & $7.8 \pm 0.3$ & $9.5 \pm 0.3$\\
$20$ & $10$ & $0.282 \pm 0.002$ & $33.7 \pm 0.8$ & $9.6 \pm 0.2$\\
$\bf 25$  & - & $\bf 1.2112 \pm 0.0004$ & $\bf 7.70 \pm 0.09$ & $\bf 9.3 \pm 0.1$ \\
  &  & & &  $\bf 9.4 \pm 0.2$\\
\hline
%\multicolumn{5}{l}{}\\
%\multicolumn{5}{p{13cm}}{}\\
\end{tabular}
\end{center}
\caption{Results for
$f={\overline{\Phi}}_{A,1}/{\overline{h}}_{\mathcal{A}}$,
$P(\lambda_B|\lambda_1)$ and $k_{AB}$, for the general and exclusive
switches. FFS results are averaged over 10 independent runs, using $n$
surfaces, as described in the text. Brute-force results, in {\bf{bold
type}}, are averaged over runs of length $3 \times 10^{8} k^{-1}$
(general) and $9\times 10^{9} k^{-1}$ (exclusive). The upper value of
$k_{AB}$ is calculated using equation (\ref{eq6}) and the lower value
using the fitting of $F(t)$ described in the text.\label{tab2}}

\end{table}

\begin{table}[b]
\begin{center}
\begin{tabular}{ccc|cc}
\hline 
\multicolumn{3}{c}{FFS}\vline &\multicolumn{2}{c}{Brute-Force}\\
\hline
 & $\lambda_B$ & CPU  &   & CPU\\
general & $20$ & $1.9$ & general & $6.8$\\
general & $25$ & $1.1$ &&\\
general & $30$ & $1.3$&&\\
exclusive & $20$ &  $2.1$& exclusive  & $90.2$\\
exclusive & $25$ & $1$&&\\
exclusive & $30$ & $1.7$&&\\
\hline
%\multicolumn{5}{p{7.5cm}}{}\\
%\multicolumn{5}{p{7.5cm}}{}\\
\end{tabular}
\end{center}
\caption{\label{tab3}
Relative CPU time required to calculate the values of $k_{AB}$ given in table \ref{tab2}, using  parameter values as in the text.  For the FFS calculations, the calculation of the flux $f={\overline{\Phi}}_{A,1}/{\overline{h}}_{\mathcal{A}}$ accounted for $20-40$\% of the total CPU time.}
\end{table}

 Table \ref{tab2} lists the results for $f={\overline{\Phi}}_{A,1}/{\overline{h}}_{\mathcal{A}}$ and  $P(\lambda_B|\lambda_1)$, as well as $k_{AB}$, for various choices of $\lambda_A$ and $\lambda_B$. The number $n$ of surfaces used in each calculation is also listed.  In all cases, the rate constant $k_{AB}$ is in good agreement with the brute-force simulation  result. An additional test is provided by measuring $f$ and $P(\lambda|\lambda_1)$ using the brute-force runs, to provide a second brute-force estimate for $k_{AB}$. This value is also listed in table \ref{tab2}  and, as expected, is in good agreement both with the value obtained by the fitting to Poisson statistics, and with the FFS results.

Table \ref{tab3} shows the relative CPU time required for each of the calculations. Even for the general switch with a relatively fast flipping rate, estimates of $k_{AB}$ are obtained by the FFS method $3-6$ times faster than using brute-force simulations, with similar accuracy. In the case of the exclusive switch, the FFS method is $40-90$ times more efficient. Table  \ref{tab3} also demonstrates that the CPU time required for an FFS calculation does not increase as the rate $k_{AB}$ decreases, in contrast to the time for brute-force calculations. Thus FFS should allow the calculation of rates of even very rare flipping events within reasonable computational time. As an example, we have calculated the rate of flipping for a more stable version of the exclusive switch, in which the rate constant for protein degradation is reduced to $0.175k$. Using the FFS method, we obtain  a rate $k_{AB} = (1.92 \pm 0.09) \times 10^{-9}k$, 500 times slower than the switch considered above. This result would have been extremely difficult to obtain using brute-force simulation.

\begin{figure}[t]
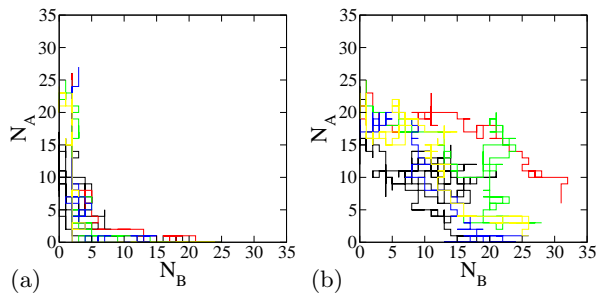

\begin{center}
\makebox[0pt][l]{(a)}{\rotatebox{0}{{\includegraphics[scale=0.2,clip=true]{fig5a}}}}\hspace{0.2cm}\makebox[0pt][l]{(b)}{\rotatebox{0}{{\includegraphics[scale=0.2,clip=true]{fig5b}}}}
\caption{Five randomly chosen transition paths, plotted as a function of $\mathrm{N_A}$ and $\mathrm{N_B}$ (a): general switch (b): exclusive switch.\label{fig5}
}
\end{center}
\end{figure}

\begin{figure}[t]
\begin{center}
\includegraphics[width=8.5cm]{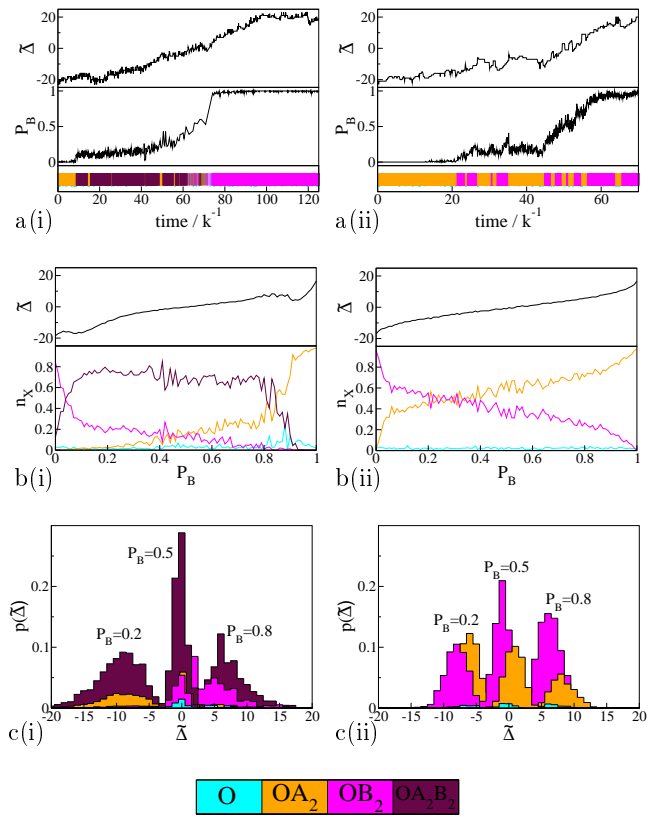}

\caption{(a): Typical transition paths, for the general (i) and
exclusive (ii) switches. $\tilde{\Delta} = (n_{\mathrm{B}} +
2n_{\mathrm{B}_2}) - (n_{\mathrm{A}} + 2n_{\mathrm{A}_2})$, $P_B$ and
operator occupancy (color-coded) are plotted versus time. The path for
the general switch (a) happens to be longer than that for the
exclusive switch (b), but this is not true of all paths in the
ensemble. (b): Average values of ${\tilde{\Delta}}$ and average
operator occupancies, as functions of the committor $P_B$, for the
general (i) and exclusive (ii) switches. (c): Probability distribution
$p(\tilde{\Delta})$, for selected values of $P_B$.
$p(\tilde{\Delta})$ is divided into color-coded contributions from the
different operator states. The sum of the areas of the histograms for
a particular $P_B$ is unity.  For all plots, $100$ test trajectories
were used to estimate $P_B$.\label{fig6}
}
\end{center}
\end{figure}

The results presented in table \ref{tab2} show that for the same mean
number of protein molecules, the exclusive switch has a flipping rate
approximately $50$ times slower than that of the general switch, in
agreement with previous work \cite{warren1}. In order to elucidate the
origin of this difference, we have analysed the flipping
mechanism. The FFS method generates a collection of switching
trajectories (members of the TPE).  Figure \ref{fig5} shows a sample
of five of these transition paths, for the general switch (a) and the exclusive
switch (b), plotted as a function of $\mathrm{N_A}$ and
$\mathrm{N_B}$. To obtain these paths, we begin with the collection of
partial trajectories that reach $\lambda_B$ from $\lambda_n$ and trace
these back via the intervening surfaces to $\lambda_A$. It is clear
from figure \ref{fig5} that in the general switch, protein
$\mathrm{A}$ is lost before protein $\mathrm{B}$ is gained, so that
the transition passes through a region of phase space where both
$\mathrm{N_A}$ and $\mathrm{N_B}$ are low. However, this is not the
case for the exclusive switch. An important quantity associated with
transition paths is the committor, $P_B(X)$. This is the probability
that a new simulation trajectory fired from point $x$ will reach
region $B$ before $A$ \cite{tps}. We have measured $P_B(x)$ for points
along the trajectories in the transition path ensembles generated
using FFS, for the general and exclusive switches. Figure
\ref{fig6}(a) shows $P_B$, as well as $\tilde{\Delta} =
(n_{\mathrm{B}} + 2n_{\mathrm{B}_2}) - (n_{\mathrm{A}} +
2n_{\mathrm{A}_2})$ and the occupancy of the operator sites, as
functions of time for typical transition paths. ${\tilde{\Delta}}$
measures the difference in the number of {\em{free}} protein
molecules: it is similar but not identical to $\Delta$.  A key point
is that for the general switch (figure \ref{fig6}a(i)), the operator
makes two important changes of state, from ${\mathrm{OA}}_2$ to
${\mathrm{OA}}_2{\mathrm{B}}_2$ early in the transition process, and
later from ${\mathrm{OA}}_2{\mathrm{B}}_2$ to ${\mathrm{OB}}_2$. Both
of these changes influence $P_B$. For the exclusive switch (figure
\ref{fig6}a(ii)), however, the operator is intermittently in states
${\mathrm{OA}}_2$ and ${\mathrm{OB}}_2$ during the transition.

In figure \ref{fig6} (b), values of ${\tilde {\Delta}}$ and operator
occupancies are shown, averaged over the paths in the TPE, as a
function of the committor $P_B$. Results are shown for the general (i)
and exclusive (ii) switches. Figure \ref{fig6} (c) analyses the state
points along the paths in the TPE which have values of $P_B=0.2$,
$P_B=0.5$ and $P_B=0.8$. For each of these values of the committor,
points are grouped according to their operator state. For each $P_B$
and operator state, the histograms in figure \ref{fig6}(c) show the
probability distribution function $p({\tilde{\Delta}})$. Clearly, for
points on a constant $P_B$ surface, the operator state and the number
of free molecules are correlated: when $\mathrm{A}$ is bound to the
DNA, on average, a larger $\tilde{\Delta}$ is required to
obtain a particular value of $P_B$ than when $\mathrm{B}$ is
bound. This shows that $P_B$, and hence the reaction co-ordinate,
depends not only upon the difference in copy number $\tilde{\Delta}$,
but also on the state of the operator. Importantly, the plots of
figure \ref{fig6} (b) and (c) are not symmetric on making the
transformations $P_B \to 1-P_B$, ${\tilde{\Delta}} \to
-{\tilde{\Delta}}$ and $\mathrm{O}\mathrm{A}_2 \leftrightarrow
\mathrm{O}\mathrm{B}_2$.  This demonstrates that the distribution of
transition paths does not follow the steady state phase space density,
which is symmetric on interchanging $\mathrm{A}$ and $\mathrm{B}$,
since the switch is by construction symmetric \cite{tenwolde}. It also
means that the TPE for the reverse transition, from $B$ to $A$, would
occupy a different region of phase space, as compared to the TPE for
the transition from $A$ to $B$ (see also supplementary material). The origin of this asymmetry is that
the dynamics of our system involves irreversible reactions (Reactions
(\ref{equationd}-\ref{equationf})). Indeed, our system does not
satisfy microscopic reversibility. Finally, figure \ref{fig6} clearly
demonstrates why the elimination of the operator state
$\mathrm{O}\mathrm{A}_2\mathrm{B}_2$ for the exclusive switches
enhances its stability with respect to that of the general switch.  In
the general switch, as soon as a $\mathrm{B}_2$ dimer is produced by
some rare fluctuation, it can bind to the DNA, switch off the
production of $\mathrm{A}$ and thereby accelerate the flipping of the
switch. For the exclusive switch, however, any $\mathrm{B}_2$ dimer
that is produced must wait for a second fluctuation by which
$\mathrm{A}_2$ is released from the DNA, before it can bind. This is
the origin of the enhanced stability of the exclusive switch.

\section*{Discussion}
This article presents the Forward Flux Sampling (FFS) method for the
calculation of the rates of rare events in stochastic kinetic
simulation schemes such as those used for biochemical reaction
networks. In contrast to previously developed methods for sampling the
transition path ensemble \cite{tps,vanerp}, FFS does not require
knowledge of the phase space density. It is this feature that makes it possible
to study rare events in biochemical networks.  Our
algorithm samples the TPE in a way that is, to our knowledge, new:
many paths are grown simultaneously from state $A$ to state $B$ in a
series of layers of partial paths, each layer forming the basis for
the next. The phase space separating stable states $A$ and $B$ is
traversed by the algorithm in a ``ratchet-like'' manner, making the
method highly suitable for very rare events, where one-at-a-time path
generation tends to be inefficient.  FFS is not applicable to systems whose dynamics is deterministic. It could be used, however,  in combination with any stochastic simulation technique. This will make it useful for a wide range of problems in soft condensed matter systems, including rare events and non-equilibrium processes.

We have demonstrated our method using stochastic simulations of a
simple genetic switch consisting of two mutually repressing
genes. Following earlier work \cite{warren1}, we compare the case
where both protein products can bind simultaneously as dimers to the
DNA (the general switch), to that where each protein dimer excludes
the binding of the other (the exclusive switch). The results obtained
using FFS are in good agreement with those of long brute-force
simulations for both switches.  The computational time required for
the FFS calculations is far less than for the brute-force simulations,
and in addition, does not increase as the rate constant
decreases. Indeed, using FFS we could simulate a switch that was too
stable to be studied using brute-force calculations. By analysing the
transition path ensembles we were able to discover the differences
between the flipping mechanisms for the general and exclusive
switches. These allow us to understand the origin of the enhanced
stability of the exclusive switch.  The FFS method will be easily
applicable to many important biochemical switches, for which
prediction of the rates and pathways of switching should lead to a
better understanding of the design principles underlying their
stability.

We thank Peter Bolhuis for very helpful discussions and Daan Frenkel,
Rutger Hermsen and Harald Tepper for their careful reading of the
manuscript. The work is part of the research program of the
``Stichting voor Fundamenteel Onderzoek der Materie (FOM)", which is
financially supported by the ``Nederlandse organisatie voor
Wetenschappelijk Onderzoek (NWO)".

\clearpage
\newpage

\begin{figure*}[t]
\begin{center}
\includegraphics[width=16cm]{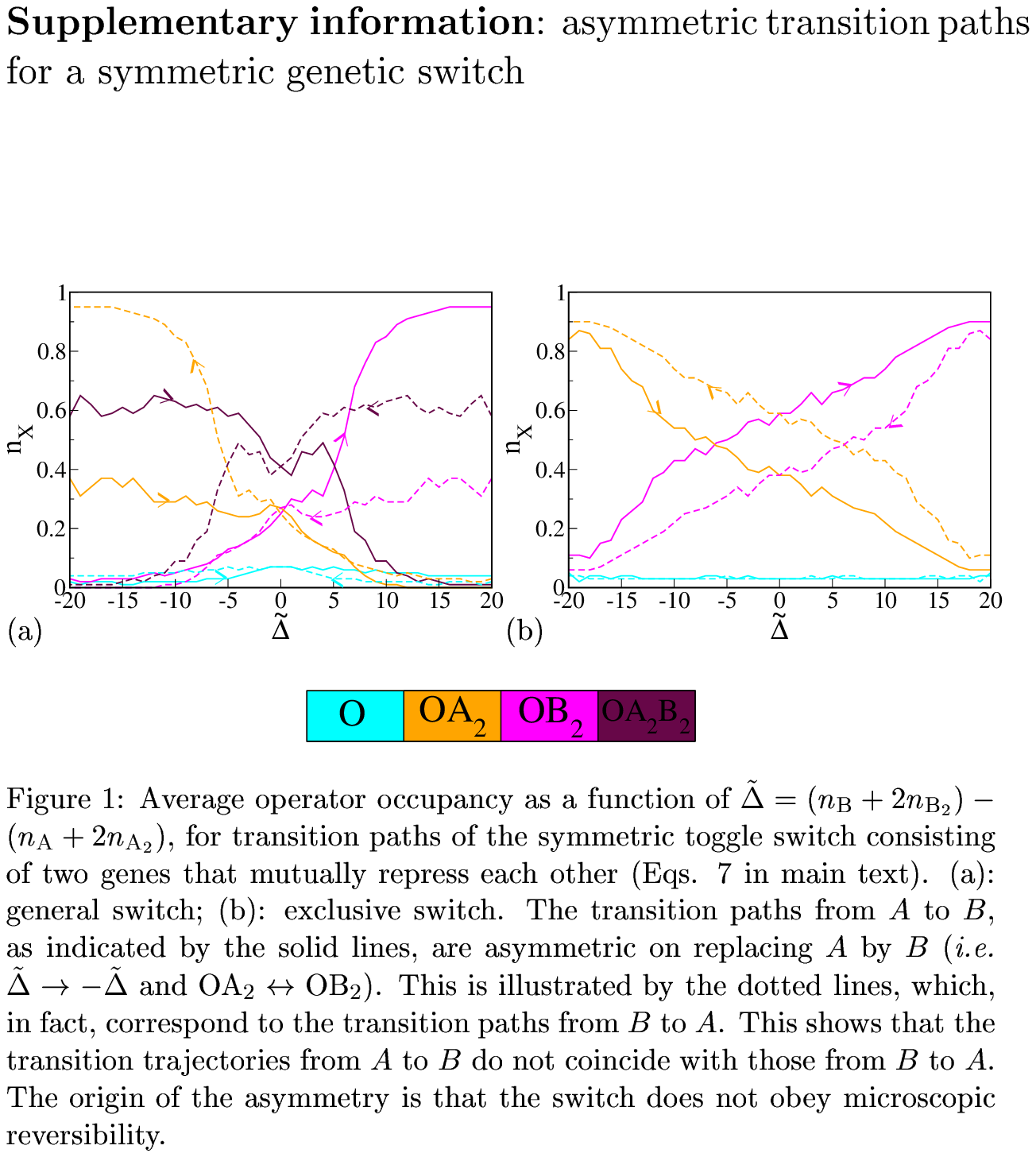}
\end{center}
\end{figure*}

\end{document}